\documentstyle[12pt]{article}
\begin{document}
\tolerance=5000
\def\pp{{\, \mid \hskip -1.5mm =}}
\def\cL{{\cal L}}
\def\be{\begin{equation}}
\def\ee{\end{equation}}
\def\bea{\begin{eqnarray}}
\def\eea{\end{eqnarray}}
\def\tr{{\rm tr}\, }
\def\nn{\nonumber \\}
\def\e{{\rm e}}
\def\D{{D \hskip -3mm /\,}}

\def\SEH{S_{\rm EH}}
\def\SGH{S_{\rm GH}}
\def\AdS5{{{\rm AdS}_5}}
\def\S4{{{\rm S}_4}}
\def\gfv{{g_{(5)}}}
\def\gfr{{g_{(4)}}}
\def\SC{{S_{\rm C}}}
\def\RH{{R_{\rm H}}}

\def\wlBox{\mbox{
\raisebox{0.3cm}{$\widetilde{\mbox{\raisebox{-0.3cm}{$\Box$}}}$}}}

\begin{center}
{\large\bf
AdS/CFT correspondence and quantum induced  
dilatonic  multi-brane-worlds
}

\vfill

{\sc Shin'ichi NOJIRI}\footnote{email: nojiri@cc.nda.ac.jp} 
and {\sc Sergei D. ODINTSOV}$^{\spadesuit,\clubsuit}$\footnote{
email: odintsov@ifug5.ugto.mx, odintsov@itp.uni-leipzig.de}

\vfill

{\sl Department of Applied Physics \\
National Defence Academy, 
Hashirimizu Yokosuka 239, JAPAN}

\vfill

{\sl $\spadesuit$ 
Tomsk State Pedagogical University,
634041 Tomsk,RUSSIA}

\vfill

{\sl $\clubsuit$ Instituto de Fisica de la Universidad de 
Guanajuato \\
Apdo.Postal E-143, 37150 Leon, Gto., MEXICO} 

\vfill

{\bf ABSTRACT}

\end{center}

d5 dilatonic gravity action with surface counterterms motivated by 
AdS/CFT correspondence and with contributions of brane quantum CFTs is 
considered around AdS-like bulk. The effective equations of motion are 
constructed. They admit two (outer and inner) or multi-brane solutions 
where brane CFTs may be different. The role of quantum brane CFT is in 
inducing of complicated brane dilatonic gravity. For exponential bulk 
potentials the number of AdS-like bulk spaces is found in analytical form.
The correspondent flat or curved (de Sitter or hyperbolic) dilatonic 
two branes are created, as a rule, thanks to quantum effects. The 
observable early Universe may correspond to inflationary brane. The found 
dilatonic quantum two brane-worlds usually contain the naked singularity 
but in couple explicit examples the curvature is finite and horizon 
(corresponding to wormhole-like space) appears.

\vfill 


\newpage

\section{Introduction}

Recent booming activity in the study of brane-worlds is caused 
by several reasons. First, gravity on 4d brane embedded in 
higher dimensional AdS-like Universe may be localized \cite{RS1,RS2}. 
Second, the way to resolve the mass hierarchy problem 
appears\cite{RS1}. Third, the new ideas on 
cosmological constant problem solution come to game \cite{ADKS,ADDK}.
Very incomplete list of references\cite{CH,cosm} (and references 
therein) mainly on the cosmological aspects of brane-worlds 
is growing every day.

The essential element of brane-world models is the presence 
in the theory of two free parameters (bulk cosmological constant 
and brane tension, or brane cosmological constant). The role of 
brane cosmological constant is to fix the position of the brane 
in terms of tension (that is why brane cosmological constant 
and brane tension are almost the same thing). 
Being completely consistent and mathematically reasonable, 
such way of doing things may look not completely satisfactory. 
Indeed, the physical 
origin (and prediction) of brane tension in terms of some dynamical 
mechanism may be required.

The ideology may be different, in the spirit of refs.\cite{NOZ, HHR}.
One considers the addition of surface counterterms to the 
original action on AdS-like space. These terms are responcible 
for making the variational  
procedure to be well-defined (in Gibbons-Hawking spirit) and for 
elimination of the leading divergences of the action. 
Brane tension is 
not considered as free parameter anymore but it is fixed by the 
condition of finiteness of spacetime when brane goes to 
infinity. Of course, leaving 
the theory in such form would rule out the possibility of consistent 
brane-world solutions existance. Fortunately, 
other parameters contribute  
to brane tension. If one considers that there is quantum 
CFT living on 
the brane (which is more close to the spirit of AdS/CFT 
correspondence\cite{AdS} ) then such CFT produces conformal anomaly 
(or anomaly induced effective action). This contributes to 
brane tension.
As a result dynamical mechanism to get brane-world with flat or 
curved (de Sitter or Anti-de Sitter) brane appears. 
The curvature of such 4d Universe is 
expressed in terms of some dimensional parameter $l$ which 
usually appears in AdS/CFT set-up and of content of quantum 
brane matter. In other words,
brane-world is the consequence of the fact (verified 
experimentally by everybody life) of the presence of 
matter on the brane! For example, sign of conformal anomaly 
terms for usual matter is such that in 
one-brane case the de Sitter (ever expanding, inflationary ) 
Universe is preferrable solution of brane 
equation\footnote{Similar mechanism for 
anomaly driven inflation in usual 4d world has been invented by 
Starobinsky\cite{SMM} and generalized for dilaton presence in 
refs.\cite{Brevik}}. 

The scenario of refs.\cite{NOZ,HHR} may be extended to the presence
of dilaton(s) as it was done in ref.\cite{NOO} or to formulation of 
quantum cosmology in Wheeler-De Witt form \cite{ANO}. 
Then whole scenario looks even more related with AdS/CFT 
correspondence as dilatonic gravity naturally follows as bosonic 
sector of d5 gauged supergravity. Moreover, the extra prize-in 
form of dynamical determination of 4d boundary value 
of dilaton-appears. In ref.\cite{NOO} the quantum dilatonic one brane 
Universe has been presented with possibility to get inflationary or 
hyperbolic or flat brane with dynamical determination of brane 
dilaton. The interesting question is related with 
generalization of such scenario in dilatonic gravity for 
multi-brane case. This will be the purpose of present work.

In the next section we present general action of d5 dilatonic 
gravity with surface counterterms and quantum brane CFT 
contribution. This action is convenient for description of 
brane-worlds where bulk is AdS-like spacetime. There could 
be one or two (flat or curved) branes in the theory. As it 
was already mentioned the brane tension is fixed in our approach,
instead of it the effective brane tension is induced by 
quantum effects. 
In section three, the explicit analytical solution of bulk 
equation for number of exponential bulk potentials is presented. 
There is the possibility to have two (inner and outer) 
branes associated with each of above bulk solutions. 
It is interesting that quantum created branes can be flat, 
or de Sitter (inflationary) or hyperbolic.
The role of quantum brane matter corrections in getting of such 
branes is extremely important. Nevertheless, there are few 
particular cases where such branes appear on classical level, 
i.e. without quantum corrections. We also briefly describe 
how to get generalization of above solutions for quantum 
dilatonic multi-brane-worlds with more than two branes. 
Brief summary of results is given in final section where 
also the result of studying the character of singularities 
for proposed two-brane solutions is presented. In most cases, 
as usually occurs in AdS dilatonic gravity, the solutions 
contain the naked singularity. However, 
in couple cases the scalar curvature is finite and there is horizon.
The corresponding 4d branes may be interpreted as wormhole.

\section{Dilatonic gravity action with brane quantum corrections}

Let us present the initial action for dilatonic AdS gravity under 
consideration. The metric of (Euclidean) AdS has the following form:
\be
\label{AdS}
ds^2=dz^2 + \sum_{i,j=1}^4 g_{(4)ij} dx^i dx^j\ ,
\quad g_{(4)ij}=\e^{2\tilde A(z)}\hat g_{ij}\ .
\ee
Here $\hat g_{ij}$ is the metric of the Einstein manifold, which is
defined by $r_{ij}=k\hat g_{ij}$, where $r_{ij}$ is 
the Ricci tensor constructed with $\hat g_{ij}$ and 
$k$ is a constant. 
One can consider two copies of the regions given by $z<z_0$ and 
glue two regions  putting a brane at $z=z_0$. 
More generally, one can consider two copies of regions 
$\tilde z_0<z<z_0$ and glue the regions putting two branes 
at $z=\tilde z_0$ and $z=z_0$. 
Hereafter we call the brane at $z=\tilde z_0$ as ``inner'' brane 
and that at $z=z_0$ as ``outer'' brane. 

Let us first consider the case with only one brane at $z=z_0$ and 
start with Euclidean signature action $S$ which is 
the sum of the Einstein-Hilbert action $\SEH$ with kinetic term 
and potential $V(\phi)={12 \over l^2}+\Phi(\phi)$ for dilaton 
$\phi$, the Gibbons-Hawking surface term $\SGH$,  the surface 
counter term $S_1$ and the trace anomaly induced action 
$W$\footnote{For the introduction to anomaly induced 
effective action in curved space-time (with torsion), see
section 5.5 in \cite{BOS}. This anomaly induced action is due to 
brane CFT living on the boundary of dilatonic AdS-like space.}: 
\bea
\label{Stotal}
S&=&\SEH + \SGH + 2 S_1 + W, \\
\label{SEHi}
\SEH&=&{1 \over 16\pi G}\int d^5 x \sqrt{\gfv}\left(R_{(5)} 
 -{1 \over 2}\nabla_\mu\phi\nabla^\mu \phi 
 + {12 \over l^2}+ \Phi(\phi) \right), \\
\label{GHi}
\SGH&=&{1 \over 8\pi G}\int d^4 x \sqrt{\gfr}\nabla_\mu n^\mu, \\
\label{S1}
S_1&=& -{1 \over 16\pi G l}\int d^4 x \sqrt{\gfr}\left(
{6 \over l} + {l \over 4}\Phi(\phi)\right), \\
\label{W}
W&=& b \int d^4x \sqrt{\widetilde g}\widetilde F A \nn
&& + b' \int d^4x\sqrt{\widetilde g}
\left\{A \left[2{\wlBox}^2 
+\widetilde R_{\mu\nu}\widetilde\nabla_\mu\widetilde\nabla_\nu 
 - {4 \over 3}\widetilde R \wlBox^2 
+ {2 \over 3}(\widetilde\nabla^\mu \widetilde R)\widetilde\nabla_\mu
\right]A \right. \nn
&& \left. + \left(\widetilde G - {2 \over 3}\wlBox \widetilde R
\right)A \right\} \\
&& -{1 \over 12}\left\{b''+ {2 \over 3}(b + b')\right\}
\int d^4x \sqrt{\widetilde g} 
\left[ \widetilde R - 6\wlBox A 
 - 6 (\widetilde\nabla_\mu A)(\widetilde \nabla^\mu A)
\right]^2 \nn
&& + C \int d^4x \sqrt{\widetilde g}
A \phi \left[{\wlBox}^2 
+ 2\widetilde R_{\mu\nu}\widetilde\nabla_\mu\widetilde\nabla_\nu 
 - {2 \over 3}\widetilde R \wlBox^2 
+ {1 \over 3}(\widetilde\nabla^\mu \widetilde R)\widetilde\nabla_\mu
\right]\phi \ .\nonumber
\eea 
Here the quantities in the  5 dimensional bulk spacetime are 
specified by the suffices $_{(5)}$ and those in the boundary 4 
dimensional spacetime are specified by $_{(4)}$. 
The factor $2$ in front of $S_1$ in (\ref{Stotal}) is coming from 
that we have two bulk regions which 
are connected with each other by the brane. 
In (\ref{GHi}), $n^\mu$ is 
the unit vector normal to the boundary. In (\ref{GHi}), (\ref{S1}) 
and (\ref{W}), one chooses 
the 4 dimensional boundary metric as 
$\gfr_{\mu\nu}=\e^{2A}\tilde g_{\mu\nu}$.  
We should distinguish $A$ and $\tilde g_{\mu\nu}$ with 
$\tilde A(z)$ and $\hat g_{ij}$ in (\ref{AdS}). The metric 
$\tilde g_{ij}$ is given by 
$\tilde g_{\mu\nu}dx^\mu dx^\nu\equiv l^2\left(d \sigma^2 
+ d\Omega^2_3\right)$. We also specify the 
quantities given by $\tilde g_{\mu\nu}$ by using $\tilde{\ }$. 
$G$ ($\tilde G$) and $F$ ($\tilde F$) are the Gauss-Bonnet
invariant and the square of the Weyl tensor. 

In the effective action (\ref{W}) induced by brane quantum matter, 
with $N$ scalar, $N_{1/2}$ spinor, $N_1$ vector fields, $N_2$ 
($=0$ or $1$) gravitons and $N_{\rm HD}$ higher 
derivative conformal scalars, $b$, $b'$ and $b''$ are 
\bea
\label{bs}
&& b={N +6N_{1/2}+12N_1 + 611 N_2 - 8N_{\rm HD} 
\over 120(4\pi)^2}\nn 
&& b'=-{N+11N_{1/2}+62N_1 + 1411 N_2 -28 N_{\rm HD} 
\over 360(4\pi)^2}\ , \quad b''=0\ .
\eea
Usually, $b''$ may be changed by the finite 
renormalization of local counterterm in gravitational 
effective action. As it was the case in ref.\cite{NOO}, the term 
proportional 
to $\left\{b''+ {2 \over 3}(b + b')\right\}$ in (\ref{W}), and 
therefore $b''$, does not contribute to the equations of motion.
Note that CFT matter induced effective action may be considered as
brane dilatonic gravity.
 
For typical examples motivated by AdS/CFT correspondence\cite{AdS} 
one has: a) ${\cal N}=4$ $SU(N)$ SYM theory: 
$b=-b'={C \over 4}={N^2 -1 \over 4(4\pi )^2}$, 
b) ${\cal N}=2$ $Sp(N)$ theory: 
$b={12 N^2 + 18 N -2 \over 24(4\pi)^2}$, 
$b'=-{12 N^2 + 12 N -1 \over 24(4\pi)^2}$.
One can write the corresponding expression for dilaton coupled spinor matter
\cite{peter} which also has non-trivial (slightly different in 
form) dilatonic contribution to CA than in case of holographic 
conformal 
anomaly\cite{LT} for ${\cal N}=4$ super Yang-Mills theory.

We can also consider the case where there are two branes at 
$z=\tilde z_0$ and $z=z_0$, adding the action corresponding 
to the brane at $z=\tilde z_0$ to the action in (\ref{Stotal}). 

\section{Dilatonic quantum brane-worlds}

Let us consider the solution of field equations 
for two-branes model. First of all, one defines a 
new coordinate $z$ by $z=\int dy\sqrt{f(y)}$ 
and solves $y$ with respect to $z$. Then the warp
factor is $\e^{2\hat A(z,k)}=y(z)$. Here one assumes 
the metric of 5 dimensional spacetime as follows:
\be
\label{DP1}
ds^2=g_{(5)\mu\nu}dx^\mu dx^\nu =f(y)dy^2 
+ y\sum_{i,j=1}^4\hat g_{ij}(x^k)dx^i dx^j. 
\ee
Here $\hat g_{ij}$ is the metric of the 4 dimensional Einstein 
manifold as in (\ref{AdS}). 

Here we only summarize the obtained results (for more details,
see \cite{NOO1}). Generally the obtained 
bulk solutions have the form: 
\be
\label{assmp1b}
\phi(y)=p_1\ln \left(p_2 y\right)\ , \quad 
\Phi(\phi)=-{12 \over l^2} + c_1 \exp\left(a\phi\right) 
+ c_2\exp\left(2a\phi\right)\ .
\ee

\noindent 
Case 1

\noindent
(a) bulk solution
\be
\label{case1b}
c_1={6kp_2p_1^2 \over 3 - 2p_1^2}\ ,\quad 
c_2=0\ ,\quad a=-{1 \over p_1}\ ,\quad p_1\neq \pm \sqrt{6} \ ,
\quad f(y)={3- 2p_1^2 \over 4ky} \ .
\ee
(b) When $k\neq 0$ and $p_1^2<2$, there is an outer brane 
solution if $F_1(y_+)\geq -8b'$, 
and there is an inner brane solution if 
$F_1(y_-)\leq 8\tilde b'$. Here $F_1$ is defined by
\be
\label{F1b}
F_1(y_0)\equiv {3 \over 16\pi G}\left(
{q \over 2}y_0^{3 \over 2} - {1 \over 2l}y_0^2
 - {q^2p_1^2ly_0 \over 16}\right) 
\ee
and $y_\pm$ is given by
\be
\label{ypm}
y_\pm^{1 \over 2}
\equiv {3ql \over 8}\left(1\pm \sqrt{1 
- {4p_1^2 \over 9}}\right)\ .
\ee
(c) Solution for $k=0$
\be
\label{c1k0ib}
p_1^2\rightarrow {3 \over 2}\ ,\quad 
y_0^{1 \over 2}={3ql \over 4}\ ,\quad {ql \over 4}\ .
\ee

\noindent
Case 2

\noindent
(a) bulk solution
\be
\label{case2b}
c_1=-6kp_2\ ,\quad 
a=\pm{1 \over \sqrt{3}}\ ,\quad p_1=\mp\sqrt{3} \ ,
\quad f(y)={3 \over {2c_2 \over p_2^2} - 4ky} \ . 
\ee
(b) In case of $k>0$, $\tilde c_2\equiv {c_2 \over p_2^2}$ 
should be positive and there is an outer brane solution, at 
least if  $F_2\left({\tilde c_2 \over 2k}\right)\geq -8b'$, where 
\be
\label{F2bbb}
F_2(y_0)\equiv {3 \over 16\pi G}\left(
{y_0 \over 2}\sqrt{2\tilde c_2 - 4ky_0 \over 3}
 - {y_0^2 \over 2l} + {kly_0 \over 4}
 - {l \tilde c_2 \over 24}\right) \ .
\ee
(c) In case of $k<0$, $F_2(y)$ has at least one mimimum 
if $\tilde c_2 < k^2l^2 \left( \sqrt{3 \over 2} -1\right)$ 
or $\tilde c_2>0$. If the value of $F_2(y)$ at the maximum is 
larger than $-8b'$, there is an outer brane solution. If 
$\tilde c_2>0$ and $F_2(0)<8\tilde b'$ or 
$\tilde c_2<0$ and $F_2\left({\tilde c_2 \over 2k}\right)
<8\tilde b'$, 
there can be an inner brane solution. 

\noindent
(d) In case of $k=0$, if $\tilde c_2>0$, 
the solution is given by
\be
\label{F20bb}
y_0={l\sqrt{\tilde c_2} \over 2}\left(\sqrt{2 \over 3} \pm 
{1 \over \sqrt{3}}\right)\ .
\ee

\noindent 
Case 3

\noindent
(a) bulk solution
\be
\label{case3b}
c_2=3kp_2\ ,\quad 
a=\pm{1 \over \sqrt{3}}\ ,\quad p_1=\mp{\sqrt{3} \over 2} \ , 
\quad f(y)={21\sqrt{p_2} \over 8\sqrt{y}\left(c_1y 
+ 7k\sqrt{p_2y}\right)} \ .
\ee
(b) When $\tilde c_1\equiv {c_1 \over \sqrt{p_2}}<0$, $k>0$ and 
there can be outer brane solution if 
$F_3\left({49 k^2 \over \tilde c_1^2}\right)>-8b'$, where
\bea
\label{F3bbb}
F_3(y_0) &\equiv& {3 \over 16\pi G}\left\{
{y_0 \over 2y_0}\sqrt{8\sqrt{y_0}\left(\tilde c_1 y_0 
+ 7k\sqrt{y_0}\right) \over 21} \right.\nn
&& \left. - {y_0^2 \over 2l} - {l \over 24}
\sqrt{y_0}\left(\tilde c_1 y_0 + 3k\sqrt{y_0}\right)
\right\}\ .
\eea
(c) When $\tilde c_1>0$ and $k<0$, 
there always exists outer brane solution if 
$F_3\left({49 k^2 \over \tilde c_1^2}\right)>-8b'$.

%%%%%
 From the above results in case 1$\sim$3, we find there very often 
appear two (inner and outer) branes solution as in the first model 
by Randall and Sundrum \cite{RS1}. Moreover, the branes may be curved 
as de Sitter or hyperbolic space which gives the  way for ever 
expanding inflationary Universe. Such 
solutions often can exist even if there is no any quantum effect, i.e., 
$b'=0$. 

Let us make few remarks on the form of metric.
If one considers the metric in the form (\ref{AdS}), 
 the warp factor $\e^{2\tilde A(z)}$ does not behave as 
an exponential function of $z$ but as a power of $z$. 
This would require that we need a region (of complete spacetime) 
where, the potential and the dilaton become almost constant. 
It results in difficulties when one tries to explain the hierarchy 
using this model. 

Hence, we presented number of dilatonic (inflationary, flat or 
hyperbolic) two brane-world Universes which are created by 
quantum effects of brane matter. Sometimes, such Universes may be 
realized due to specific choice of dilatonic potential even 
on classical level.

In some papers (for example in \cite{HSTT}), the solution with 
many branes was proposed. In such model, there are two AdS 
spaces with the different radii or different values of the 
cosmological constants. They are glued by a brane, whose 
tension is given by the difference of the inverse
of the radii. In the solution, the value of ${dA \over dz}$ in 
the metric of the form (\ref{AdS}) jumps at the brane, which 
tells the value of $f(y)$ in the metric choice in (\ref{DP1}) 
jumps on the brane since $\sqrt{f(y)}={dz \over dy}
={1 \over 2y {dA \over dz}}$. 
Imagine one includes the quantum effects on the brane. Then one can, 
in general, glue  two AdS-like spaces with same values of 
the cosmological constant. Let us assume that there is a brane at 
$y=\hat y_0$ and there are two AdS-like spaces in $y>\hat y_0$ 
and $y<\hat y_0$ glued by the brane. One now denotes the quantity 
in the AdS-like space in $y>\hat y_0$ ($y<\hat y_0$) by the 
suffix $+$ ($-$). Then we can often generalize two brane-world 
for multi-brane case.

\section{Discussion} 

In summary, we presented the generalization of quantum dilatonic 
brane-world\cite{NOO} where brane is flat, spherical (de Sitter) or 
hyperbolic and it is induced by quantum effects of CFT living on 
the brane.
In this generalization one may have two brane-worlds or even 
multi-brane-worlds which proves general character of scenario 
suggested in refs.\cite{HHR,NOZ} where instead of arbitrary 
brane tension added by hands the effective brane tension is 
produced by boundary quantum fields. 
What is more interesting the bulk solutions have analytical form,
 at least, for specific 
choice of bulk potential under consideration.

In classical dilatonic gravity the variety of brane-world solutions has 
been presented in ref.\cite{CEGH} where also the question of 
singularities has been discussed. The fine-tuned example of bulk
potential where one gets bulk solution which is not singular has been 
presented. In our solutions, the curvature singularity appears 
at $y=0$. 

In case 1, when $y\sim 0$ and the coordinates besides $y$ are 
fixed, the infinitesimally small distance $ds$ is given by 
$ds=\sqrt{f}dy \sim {dy \over q\sqrt{y}}$,
which tells that the distance between the brane and the 
singularity is finite. Then in cases of $k=0$ and $k<0$, 
the singularity is naked when we Wick re-rotate spacetime to 
Lorentzian signature. When $k>0$, the singularity is not 
exactly naked after the Wick re-rotation since the horizon 
is given by $y=0$, i.e. the horizon coincides with the 
curvature singularity. 

In case 2, the situation is not changed for $k=0$, $k>0$ 
 and $k<0$ with $\tilde c_2>0$ from that in case 1 and 
the distance between the brane and the singularity is finite 
since $ds\sim {dy \over \sqrt{y}}
\sqrt{3 \over 2\tilde c_2}$ when $y$ is small. 
When   $k<0$ with $\tilde c_2<0$, however, 
the singularity is not naked since there is a kind of horizon 
at $y={\tilde c_2 \over 2k}$, where ${1 \over f(y)}=0$. 
We should note the scalar curvature $R_{(5)}$ 
is finite. This tells that $y$ is not proper coordinate when 
$y\sim{\tilde c_2 \over 2k}$. If  new coordinate $\eta$ is 
introduced :
$\eta^2\equiv 2\left(y - {\tilde c_2 \over 2k}\right)$, 
the metric in (\ref{DP1}) is rewritten as follows,
\be
\label{DP1b}
ds^2=-{3 \over 4k}d\eta^2 + \left({\tilde c_2 \over 2k}
+ {\eta^2 \over 2}\right)
\sum_{i,j=1}^4\hat g_{ij}(x^k)dx^i dx^j\ . 
\ee
The radius of 4d manifold with negative $k$, whose metric is 
given by $\hat g_{ij}$, has a minimum ${\tilde c_2 \over 2k}$ 
at $\eta=0$, which 
corresponds to $y={\tilde c_2 \over 2k}$. The radius increases 
when $|\eta|$ increases. Therefore the spacetime can be regarded 
as a kind of wormhole, where two universes corresponding to 
$\eta>0$ and $\eta<0$, respectively, are joined at $\eta=0$. 

In case 3, the singularity is naked (the singularity 
is not exactly naked when $k>0$ as in case 1) in general 
and the distance between the brane and the horizon is finite 
except $k>0$ and $\tilde c_1<0$ case since there is a horizon 
at $\sqrt{y}=-{7k \over \tilde c_1}$ where the scalar curvature 
is finite.
 
The price for having analytical bulk results (exactly solvable bulk 
potential) is the presence of (naked) singularity. One can, of course,
present the fine-tuned examples of bulk potential as in refs.\cite{NOO,CEGH}
where the problem of singularity does not appear. Moreover, bulk quantum 
effects may significally modify classical bulk configurations \cite{NOZ,NOZ2,
GPT} which presumbly may help in resolution of (naked) singularity problem.
 However, in such situation there are no analytical bulk solutions in
dilatonic gravity.

There are various ways to extend the results of present work.
First of all, one can construct multi-brane dilatonic solutions within 
the current scenario for another classes of bulk potential. However, this 
requires the application of numerical methods.
Second, it would be interesting to describe the details of brane-world 
anomaly driven inflation (with non-trivial dilaton) at late times when 
it should decay to standard FRW cosmology.
Third, within similar scenario 
one can consider dilatonic brane-world black holes which  
are currently under investigation.

\

\noindent{\bf Acknowledgements}
SDO would like to thank organisers of Fourth Mexican School: Membranes 2000
 especially O. Obregon and J. Socorro for kind possibility to present
 invited talk 
at the School.


\begin{thebibliography}{99}
\bibitem{RS1} L. Randall and R. Sundrum,
 {\sl Phys.Rev.Lett.} {\bf 83} (1999) 3370, hep-th/9905221.
\bibitem{RS2} L. Randall and R. Sundrum,
 {\sl Phys.Rev.Lett.} 
 {\bf 83} (1999)4690, hep-th/9906064. 
\bibitem{CH} A. Chamblin and H.S. Reall, hep-th/9903225, 
{\sl Nucl.Phys.} {\bf B562} (1999) 133;
N. Kaloper, hep-th/9905210, 
{\sl Phys.Rev.} {\bf D60} (1999) 123506;
A. Lukas, B. Ovrut and D. Waldram, hep-th/9902071, 
{\sl Phys.Rev.} {\bf D61} (2000) 064003;
T. Nihei, hep-th/9905487, {\sl Phys.Lett.} {\bf B465} (1999) 81;
H. Kim and H. Kim, hep-th/9909053, {\sl Phys.Rev.} 
{\bf D61} (2000) 064003;
D. Chung and K. Freese, {\sl Phys.Rev.} {\bf D61} (2000) 023511; 
J. Garriga and M. Sasaki, hep-th/9912118;
K. Koyama and J. Soda, hep-th/9912118;
J. Kim and B. Kyae, hep-th/0005139, {\sl Phys.Lett.} 
{\bf B486} (2000) 165;
R. Maartens, D. Wands, B. Bassett and T. Heard,hep-ph/9912464;
S. Mukohyama, hep-th/0007239;
L. Mendes and A. Mazumdar, gr-qc/0009017.
\bibitem{cosm}
P. Binetruy, C. Deffayet and D. Langlois, hep-th/9905012, 
{\sl Nucl.Phys.} {\bf B565} (2000) 269;
J. Cline, C. Grojean and G. Servant, {\sl Phys.Rev.Lett.}
{\bf 83} (1999) 4245;
S. Giddings, E. Katz and L. Randall, 
{\sl JHEP} {\bf 0003} (2000) 023;
E. Flanagan, S. Tye and I. Wasserman, hep-ph/9909373;
C. Csaki, M. Graesser, C. Kolda and J. Terning,
{\sl Phys.Lett.} {\bf B462} (1999) 34;
P. Kanti, I. Kogan, K. Olive and M. Pospelov, {\sl Phys.Lett.}
{\bf B468} (1999) 31;
S.Mukohyama, T. Shiromizu and K. Maeda, hep-th/9912287, 
{\sl Phys.Rev.} {\bf D62} (2000) 024028;
K. Behrndt and M. Cvetic, hep-th/9909058;
M. Cvetic and J. Wang, hep-th/9912187;
R. Kallosh and A. Linde, {\sl JHEP} {\bf 0002} (2000) 005;
D. Youm, hep-th/0002147;
J. Chen, M. Luty and E. Ponton, hep-th/0003067;
S. de Alwis, A. Flournoy and N. Irges, hep-th/0004125;
R. Gregory, V.A. Rubakov and S. Sibiryakov,{\sl Phys.Rev.Lett.}
 {\bf 84} (2000)5928;
S. Nojiri and S.D. Odintsov, hep-th/0006232, 
{\sl JHEP} {\bf 0007}(2000) 049;
C. Zhu, hep-th/0005230, {\sl JHEP} {\bf 0006} (2000) 034;
H. Davouddiasl, J. Hewett and T. Rizzo, hep-ph/0006041;
P. Binetruy, J.M. Cline and C. Crojean, hep-th/0007029;
N. Mavromatos and J. Rizos, hep-th/0008074;
I. Neupane, hep-th/0008190;
K. Akama and T. Hattori, hep-th/0008133;
C. Barcelo and M. Visser, gr-qc/0008008;
\bibitem{HHR} S.W. Hawking, T. Hertog and H.S. Reall, 
hep-th/0003052, {\sl Phys.Rev.} {\bf D62} (2000) 043501.
\bibitem{NOZ} S. Nojiri, S.D. Odintsov and S. Zerbini,
hep-th/0001192, {\sl Phys.Rev.} {\bf D62} (2000)064006;
 S. Nojiri and S.D. Odintsov,
 hep-th/0004097, {\sl Phys.Lett.} {\bf B484} (2000)119.
\bibitem{ADKS} N. Arkani-Hamed, S. Dimopoulos, N. Kaloper and 
R. Sundrum, hep-th/0001197;
S. Kachru, M. Schultz and E. Silverstein, hep-th/0001206.
%%%%%%%%%%%%%%%%%%%%%%%%%%%%%%%%%%%%%%%%%%%%%%%%%%%%%%%%%%%%
\bibitem{ADDK}
N. Arkani-Hamed, S. Dimopoulos, G. Dvali and N. Kaloper, 
 hep-th/9907209, {\sl Phys.Rev.Lett.} {\bf 84} (2000) 586.

\bibitem{AdS} J.M. Maldacena, {\sl Adv.Theor.Math.Phys.} {\bf 2}
(1998)231;
E. Witten, {\sl Adv.Theor.Math.Phys.}{\bf 2} (1998)253;
S. Gubser, I.R. Klebanov and A.M. Polyakov,{\sl Phys.Lett.}{\bf B428}
(1998)105.
%%%%%%%%%%%%%%%%%%%%%%%%%%%%%%%%%%%%%%%%%%%%%%%%%%%%%%%%%%%%
%%%%%%%%%%%%%%%%%%%%%%%%%%%%%%%%%%%%%%%%%%%%%%%%%%%%%%%%%%%%
\bibitem{NOO} S. Nojiri, O. Obregon and S.D. Odintsov, 
hep-th/0005127, {\sl Phys.Rev.} {\bf D62} (2000) 104003.
\bibitem{ANO} L. Anchordoqui, C. Nunez and K. Olsen, hep-th/0007064;
L. Anchordoqui and K. Olsen, hep-th/0008102
\bibitem{BOS} I.L. Buchbinder, S.D. Odintsov and I.L. Shapiro,
Effective Action in Quantum Gravity,
IOP Publishing, Bristol and Philadelphia 1992.
\bibitem{LT} H. Liu and A. Tseytlin, {\sl Nucl.Phys.} {\bf B533} 
(1998) 88, hep-th/9804083;
S. Nojiri and S.D. Odintsov, {\sl Phys.Lett.} {\bf B444} (1998) 92.
\bibitem{peter} P. van Nieuwenhuizen, S. Nojiri and S.D. Odintsov,
{\sl Phys.Rev.} {\bf D 60} (1999) 084014.
\bibitem{SMM} A. Starobinsky, {\sl Phys.Lett.} {\bf B91} (1980) 99;

\bibitem{CEGH} C. Cs\'aki, J. Erlich, C. Crojean and 
T.J. Hollowood, hep-th/0004133;

\bibitem{Brevik} I. Brevik and S.D. Odintsov, {\sl Phys.Lett.} 
 {\bf B455} (1999) 104, hep-th/9902418;
B. Geyer, S.D. Odintsov and S. Zerbini, {\sl Phys.Lett.} {\bf B460} 
(1999) 58.
\bibitem{HSTT} H. Hatanaka, M. Sakamoto, M. Tachibana and 
K. Takenaga, {\sl Prog.Theor.Phys.} {\bf 102} (1999) 1213, 
hep-th/9909076. 
\bibitem{NOZ2} S. Nojiri, S.D. Odintsov and S. Zerbini, 
hep-th/0006115, {\bf Class.Quant.Grav.} {\bf 17} (2000) 4855.
\bibitem{GPT} J. Garriga, O. Pujolas and T. Tanaka, hep-th/004109.
\bibitem{NOO1} S. Nojiri,S.D. Odintsov and K.E. Osetrin, hep-th/0009059,
 {\sl Phys.Rev.D}, to appear.
\end{thebibliography}
\end{document}